\documentclass[fleqn,12pt,twoside]{article}
\usepackage[headings]{espcrc1}

\readRCS
$Id: espcrc1.tex,v 1.2 2004/02/24 11:22:11 spepping Exp $
\usepackage{graphicx}
\usepackage[figuresright]{rotating}

\newcommand{\AmS}{{\protect\the\textfont2
  A\kern-.1667em\lower.5ex\hbox{M}\kern-.125emS}}

\title{Nuclear constraints on the inner edge of neutron star crusts}

\author{Lie-Wen Chen\address[SJTU]{Department of Physics,
        Shanghai Jiao Tong University, Shanghai 200240, China}
        \thanks{Supported in part by the NNSF of China under Grant No. 10575071, 10675082 and
        10975097, Shanghai Rising-Star Program under Grant No.06QA14024, and
        the National Basic Research Program of China (973 Program) under Contract No.2007CB815004.},
        Bao-An Li\address{Department of Physics and Astronomy, Texas A\&M University-Commerce, Commerce, \\
        Texas 75429-3011, USA}
        \thanks{Supported in part by the US NSF under Grants No. PHY0652548 and No. PHY0757839, the Texas Coordinating Board of Higher Education ARP grant No. 003565-0004-2007 and the Research Corporation Grant No. 7123.},
        Hong-Ru Ma\addressmark[SJTU],
                and
        Jun Xu\addressmark[SJTU]\thanks{Current address: Cyclotron Institute and Physics Department,
        Texas A\&M University, College Station, Texas 77843-3366, USA}}

\runtitle{Nuclear constraints on the inner edge of neutron star
crusts} \runauthor{L.W. Chen, B.A. Li, H.R. Ma, and J. Xu}

\begin{document}

\maketitle

\begin{abstract}
We show that the widely used parabolic approximation to the Equation
of State (EOS) of asymmetric nuclear matter leads systematically to
significantly higher core-crust transition densities and pressures.
Using an EOS for neutron-rich nuclear matter constrained by the
isospin diffusion data from heavy-ion reactions in the same
sub-saturation density range as the neutron star crust, the density
and pressure at the inner edge separating the liquid core from the
solid crust of neutron stars are determined to be $0.040$ fm$^{-3}$
$\leq \rho _{t}\leq 0.065$ fm$^{-3}$ and $0.01$ MeV/fm$^{3}$ $\leq
P_{t}\leq 0.26$ MeV/fm$^{3}$, respectively. Implications of these
constraints on the Vela pulsar are discussed.
\end{abstract}

\section{INTRODUCTION}
The inner crust in neutron stars spans the region from the neutron
drip-out point to the inner edge separating the solid crust from the
homogeneous liquid core and plays an important role in understanding
many astrophysical observations~\cite{Pet95b,Lat00,Lin99}. While the
neutron drip-out density $\rho _{out}$ is relatively well determined
to be about $4\times 10^{11}$ g/cm$^{3}$ \cite{Rus06}, the
transition density $\rho _{t}$ at the inner edge is still largely
uncertain mainly because of our very limited knowledge on the EOS of
neutron-rich nucleonic matter, especially the density dependence of
the nuclear symmetry energy $E_{sym}(\rho)$~\cite{Lat00}.

Recently, significant progress has been made in constraining the EOS
of neutron-rich nuclear matter using terrestrial laboratory
experiments (See, e.g., Ref.~\cite{LCK08} for the most recent
review). In particular, the analysis of isospin-diffusion data
\cite{Tsa04,Che05a,LiBA05c} in heavy-ion collisions has constrained
tightly the $E_{sym}(\rho)$ in exactly the same sub-saturation
density region around the expected inner edge of neutron star crust.
In the present talk, we report our recent work on locating the inner
edge of the neutron star crust using terrestrial nuclear laboratory
data within both thermodynamical and dynamical methods~\cite{Xu09a}.

\begin{figure}[htb]
\begin{minipage}[t]{78mm}
\hspace*{-12mm}
\centering
\includegraphics[scale=0.7]{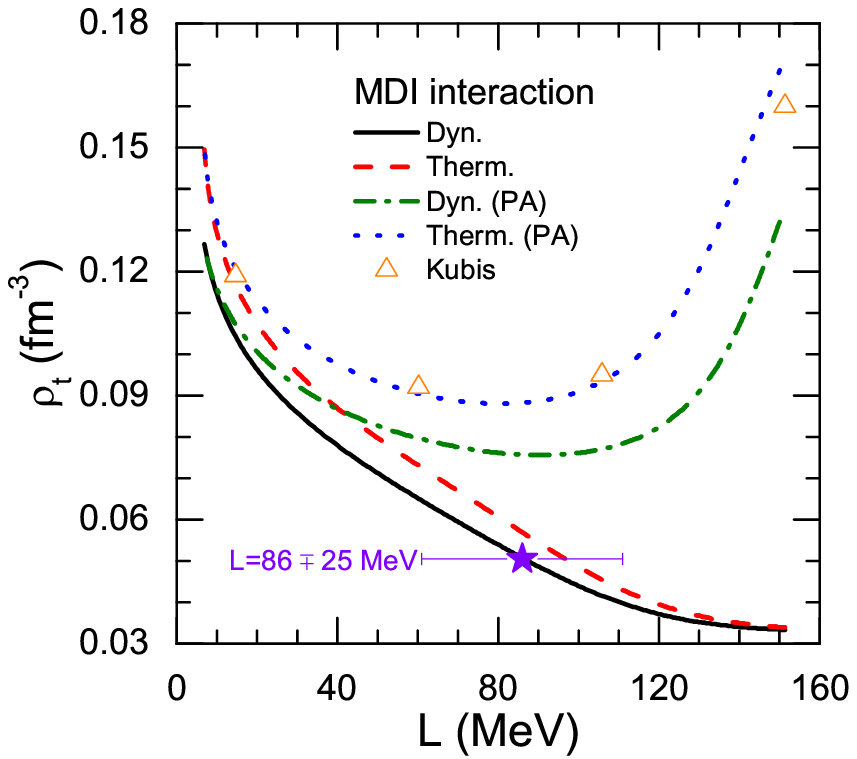}
\caption{{\protect\small (Color online) The }$\protect%
\rho _{t}$ {\protect\small verus $L$ from the dynamical and
thermodynamical methods with and without the PA using the MDI
interaction. The triangles are obtained by
Kubis~\protect\cite{Kub07} and the star with error bar represents
}${\protect\small L=86}\pm {\protect\small 25}${\protect\small \
MeV. Taken from Ref.~\cite{Xu09a}}} \label{rhotL}
\end{minipage}
\hspace{\fill}
\begin{minipage}[t]{78mm}
\hspace*{-12mm}
\centering
\includegraphics[scale=0.8]{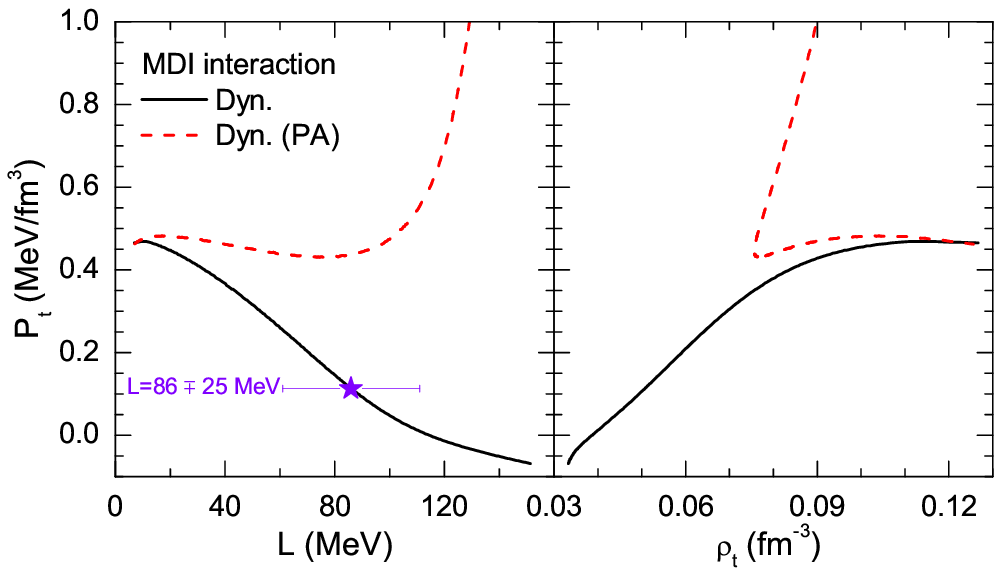}
\caption{{\protect\small (Color online) The $P_{t}$ versus $L$ and $%
\protect\rho _{t}$ by using the dynamical method with and without
the PA using the MDI interaction. The star with error bar in the
left panel represents }${\protect\small L=86}\pm {\protect\small
25}$ {\protect\small \ MeV. Taken from Ref.~\cite{Xu09a}}}
\label{PtLrhot}
\end{minipage}
\end{figure}

\section{RESULTS}

Shown in Fig.~\ref{rhotL} is the $\rho _{t}$ as a
function of the slope parameter of the symmetry energy $L=3\rho _{0}\frac{%
\partial E_{sym}(\rho )}{\partial \rho }|_{\rho =\rho _{0}}$ with the MDI
interaction~\cite{Das03}. For comparisons, we have included results
using both the dynamical and thermodynamical methods with the full
EOS and its parabolic approximation (PA), i.e., $E(\rho ,\delta
)=E(\rho ,\delta =0)+E_{sym}(\rho )\delta ^{2}+O(\delta^{4})$ using
the same MDI interaction. The thermodynamical method is the long
wave length limit of the dynamical one when the Coloumb and surface
interactions are neglected \cite{Xu09a}. With the full MDI EOS, it
is clearly seen that the $\rho _{t}$ decreases almost linearly with
increasing $L$ within both methods. It is interesting to see that
the two methods give very similar results (the difference is
actually less than $0.01$ fm$^{-3}$). Surprisingly, the PA
drastically changes the results, especially for stiffer symmetry
energies (larger $L$ values). Also included in Fig.~\ref{rhotL} are
the predictions by Kubis using the PA of the MDI EOS in the
thermodynamical approach~\cite{Kub07}. The large error introduced by
the PA is understandable since the $\beta $-stable $npe$ matter is
usually highly neutron-rich, thus the contribution from the higher
order terms in $\delta $ is appreciable. This is especially true for
the stiffer symmetry energy which generally leads to a more
neutron-rich $npe $ matter at subsaturation densities. In addition,
simply because of the energy curvatures involved in the stability
conditions, the contributions from higher order terms in the EOS are
multiplied by a larger factor than the quadratic term. Our results
indicate that one may introduce a huge error by assuming {\it a
priori} that the EOS is parabolic with respect to the isospin
asymmetry for a given interaction in calculating the $\rho _{t}$. We
thus apply the experimentally constrained $L$ to the $\rho_t-L$
correlation obtained using the full EOS in constraining the $\rho
_{t}$ in the dynamical method as it is more complete and realistic.
As shown in Fig.~\ref{rhotL}, the constrained $L=86\pm 25$ MeV
obtained from the transport model analysis of the isospin diffusion
data~\cite{Tsa04,Che05a,LiBA05c} then limits the transition density
to $0.040$ fm$^{-3} $ $\leq \rho _{t}\leq 0.065$ fm$^{-3}$.

The pressure at the inner edge, $P_{t}$, is also an important
quantity which might be measurable indirectly from observations of
pulsar glitches \cite{Lat00,Lin99}. Shown in Fig.~\ref{PtLrhot} is
the $P_{t}$ versus $L$ and $\rho _{t}$. Again, it is seen that the
PA leads to huge errors for larger (smaller) $L$ ($\rho _{t}$)
values. For the full MDI EOS, the $P_{t}$ decreases (increases) with
the increasing $L$ ($\rho _{t}$) while it displays a complex
relation with $L$ or $\rho _{t}$ for the PA. The complex behaviors
are due to the fact that the $\rho _{t}$ does not vary monotonically
with $L$ for the PA as shown in Fig.~\ref{rhotL}. From the
constrained $L$ values, the $P_{t}$ is limited between $0.01$
MeV/fm$^{3}$ and $0.26$ MeV/fm$^{3}$.

The constrained values of $\rho _{t}$ and $P_{t}$ have important
implications on many properties of neutron
stars~\cite{Pet95b,Lat00,Lin99}. As it was shown in
Ref.~\cite{Lat00}, the crustal fraction of the total moment of
inertia of a neutron star, i.e., $\Delta I/I$ depends sensitively on
the $P_t$ and $\rho_t$ at subsaturation densities, but there is no
explicit dependence upon the higher-density EOS. So far, the only
known limit of $\Delta I/I>0.014$ was extracted from studying the
glitches of the Vela pulsar \cite{Lin99}. This together with the
upper bounds on
the $P_t$ and $\rho_t$ ($\rho _{t}=0.065$ fm$^{-3}$ and $%
P_{t}=0.26$ MeV/fm$^{3}$) sets approximately a minimum radius of
$R\geq 4.7+4.0M/M_{\odot }$ km for the Vela pulsar. The radius of
the Vela pulsar is predicted to exceed $10.5$ km should it have a
mass of $1.4M_{\odot }$. A more restrictive constraint will be
obtained from the lower bounds of $\rho _{t}=0.040$ fm$^{-3}$
($P_{t}=0.01$ MeV/fm$^{3}$) and it can be approximately
parameterized by $R=5.5+14.5M/M_{\odot }$ km. It is thus seen that
the errors in both the transition density and pressure are still
large. Thus, the uncertainties for the mass-radius relation of the
Vela pulsar are still large. A conservative constraint of $R\geq
4.7+4.0M/M_{\odot }$ km using the upper bounds on the $P_t$ and
$\rho_t$ was then obtained \cite{Xu09a}. We notice that a constraint
of $R\geq 3.6+3.9M/M_{\odot }$ km for this pulsar was previously
derived in Ref.~\cite{Lin99} by using $\rho _{t}=0.075$ fm$^{-3}$
and $P_{t}=0.65$ MeV/fm$^{3}$. However, the constraint obtained
using data from both the terrestrial laboratory experiments and
astrophysical observations is significantly more stringent.

\end{document}